\documentclass[epj,nopacs]{svjour}
\usepackage{graphics}
\usepackage{amssymb} \usepackage{amsmath} 
\usepackage{graphicx}
\usepackage{epsfig,latexsym}

\begin{document}







\def\bef{\begin{figure}}
\def\eef{\end{figure}}
\newcommand{\ans}{ansatz }
\newcommand{\be}[1]{\begin{equation}\label{#1}}
\newcommand{\beq}{\begin{equation}}
\newcommand{\ee}{\end{equation}}
\newcommand{\beqn}[1]{\begin{eqnarray}\label{#1}}
\newcommand{\eeqn}{\end{eqnarray}}
\newcommand{\bd}{\begin{displaymath}}
\newcommand{\ed}{\end{displaymath}}
\newcommand{\mat}[4]{\left(\begin{array}{cc}{#1}&{#2}\\{#3}&{#4}
\end{array}\right)}
\newcommand{\matr}[9]{\left(\begin{array}{ccc}{#1}&{#2}&{#3}\\
{#4}&{#5}&{#6}\\{#7}&{#8}&{#9}\end{array}\right)}
\newcommand{\matrr}[6]{\left(\begin{array}{cc}{#1}&{#2}\\
{#3}&{#4}\\{#5}&{#6}\end{array}\right)}
\newcommand{\cvb}[3]{#1^{#2}_{#3}}
\def\lsim{\raise0.3ex\hbox{$\;<$\kern-0.75em\raise-1.1ex
e\hbox{$\sim\;$}}}
\def\gsim{\raise0.3ex\hbox{$\;>$\kern-0.75em\raise-1.1ex
\hbox{$\sim\;$}}}
\def\abs#1{\left| #1\right|}
\def\simlt{\mathrel{\lower2.5pt\vbox{\lineskip=0pt\baselineskip=0pt
           \hbox{$<$}\hbox{$\sim$}}}}
\def\simgt{\mathrel{\lower2.5pt\vbox{\lineskip=0pt\baselineskip=0pt
           \hbox{$>$}\hbox{$\sim$}}}}
\def\unity{{\hbox{1\kern-.8mm l}}}
\newcommand{\eps}{\varepsilon}
\def\ep{\epsilon}
\def\ga{\gamma}
\def\Ga{\Gamma}
\def\om{\omega}
\def\omp{{\omega^\prime}}
\def\Om{\Omega}
\def\la{\lambda}
\def\La{\Lambda}
\def\al{\alpha}
\newcommand{\ov}{\overline}
\renewcommand{\to}{\rightarrow}
\renewcommand{\vec}[1]{\mathbf{#1}}
\newcommand{\vect}[1]{\mbox{\boldmath$#1$}}
\def\tm{{\widetilde{m}}}
\def\mcirc{{\stackrel{o}{m}}}
\newcommand{\Dm}{\Delta m}
\newcommand{\dm}{\varepsilon}
\newcommand{\tanb}{\tan\beta}
\newcommand{\nbar}{\tilde{n}}
\newcommand\PM[1]{\begin{pmatrix}#1\end{pmatrix}}
\newcommand{\up}{\uparrow}
\newcommand{\down}{\downarrow}
\def\omE{\omega_{\rm Ter}}
\def\nn{\\  \nonumber}
\def\de{\partial}
\def\brf{{\mathbf f}}
\def\bbf{\bar{\bf f}}
\def\bF{{\bf F}}
\def\bbF{\bar{\bf F}}
\def\bA{{\mathbf A}}
\def\bB{{\mathbf B}}
\def\bG{{\mathbf G}}
\def\bI{{\mathbf I}}
\def\bM{{\mathbf M}}
\def\bY{{\mathbf Y}}
\def\bX{{\mathbf X}}
\def\bS{{\mathbf S}}
\def\bb{{\mathbf b}}
\def\bh{{\mathbf h}}
\def\bg{{\mathbf g}}
\def\bla{{\mathbf \la}}
\def\bmu{\mathbf m }
\def\by{{\mathbf y}}
\def\bmu{\mbox{\boldmath $\mu$} }
\def\bsig{\mbox{\boldmath $\sigma$} }
\def\bunity{{\mathbf 1}}
\def\cA{{\cal A}}
\def\cB{{\cal B}}
\def\cC{{\cal C}}
\def\cD{{\cal D}}
\def\cF{{\cal F}}
\def\cG{{\cal G}}
\def\cH{{\cal H}}
\def\cI{{\cal I}}
\def\cL{{\cal L}}
\def\cN{{\cal N}}
\def\cM{{\cal M}}
\def\cO{{\cal O}}
\def\cR{{\cal R}}
\def\cS{{\cal S}}
\def\cT{{\cal T}}
%



\def\lsim{\mathrel{\mathop  {\hbox{\lower0.5ex\hbox{$\sim$}
\kern-0.8em\lower-0.7ex\hbox{$<$}}}}}
\def\gsim{\mathrel{\mathop  {\hbox{\lower0.5ex\hbox{$\sim$}
\kern-0.8em\lower-0.7ex\hbox{$>$}}}}}

%
\title{More about neutron -- mirror neutron oscillation}
\author{Zurab Berezhiani 
}                     
%
%
\institute{Dipartimento di Fisica, Universit\`a di L'Aquila, 
I-67010 Coppito, AQ, and \\
INFN, Laboratori Nazionali del Gran Sasso,
 I-67010 Assergi, AQ, Italy }
\date{Received: date / Revised version: date}
%
\abstract{
It was pointed out recently that oscillation of 
the neutron $n$  into mirror neutron $n'$,
a sterile twin of the neutron with exactly the same mass,
could be a very fast process with the the baryon number
violation, even faster than the neutron decay itself.
This process is sensitive to the magnetic fields and
it could be observed by comparing the neutron lose rates
in the UCN storage chambers for different magnetic backgrounds.
We calculate the probability of $n-n'$ oscillation in the case  
when a mirror magnetic field $\vec{B}'$ is non-zero 
and show that in this case it can be suppressed or resonantly enhanced  
by applying the ordinary magnetic field $\vec{B}$, depending on its strength and on its orientation with respect to $\vec{B}'$. 
The recent experimental data, under this hypothesis, still allow 
the $n-n'$ oscillation time order 1 s or even smaller. 
Moreover, they indicate that the neutron losses  are sensitive to the 
orientation of the magnetic field. 
If these hints will be confirmed in the future experiments,   
this would point to the presence of the mirror magnetic field on 
the Earth of the order of  $0.1$ G,
or some equivalent  spin-dependent force of the other 
origin that makes a difference
between the neutron and mirror neutron states.
\PACS{
      {PACS-key}{discribing text of that key}   \and
      {PACS-key}{discribing text of that key}
     } 
} 
\maketitle
%


\section{Introduction}

Along with the ordinary particle world,
there may exist a hidden gauge sector in the form of its exact copy.
Such a parallel sector, coined as mirror world \cite{Mirror},
can have many interesting phenomenological and cosmological implications
(for reviews, see \cite{IJMPA-B}).
The Universe,
besides the ordinary particles: electrons, nucleons, photons, etc.,
should also contain their invisible twins: mirror electrons,
mirror nucleons, mirror photons, etc.
having exactly the same mass spectrum and coupling constants.
Mirror matter, being dark in terms of ordinary photons
and interacting with ordinary matter via gravity,
can be a viable candidate for dark matter \cite{BCV}
(for earlier works, see also \cite{Blin}).
The baryon asymmetry of the Universe can be generated
via the out-of-equilibrium,  $B\!-\!L$ and $CP$ violating processes
between the ordinary and mirror particles \cite{BB-PRL}
which mechanism naturally explains the intriguing correspondence
between the visible and dark matter fractions in the Universe
\cite{AIP}. These processes can be mediated by some heavy 
gauge singlet particles as are the right handed neutrinos \cite{BB-PRL},
or by extra heavy gauge bosons/gauginos interacting with both
sectors  \cite{PLB98}.

These interactions can also induce the  particle mixing and oscillation
phenomena between the ordinary and mirror sectors.
Any neutral particle, elementary or composite,
can have a mixing with its mirror counterpart:
e.g., photon  with mirror photon \cite{Holdom},
neutrinos  with mirror neutrinos ~\cite{FV},
pions with mirror pions \cite{IJMPA-B}, etc.

The mixing  between the ordinary neutron $n$ and its mirror partner $n'$
via a small mass term  $\dm \,(\ov{n} n' + \ov{n}' n)$
was suggested in ref. \cite{BB-nn}. It was pointed out that
the present experimental limits do not exclude
a rapid $n -n'$ oscillation,
with the timescale $\tau_{nn'}=\dm^{-1}$ of order 1 s
or at least much smaller than
the neutron decay time $\tau_{\rm dec} \simeq 10^3$~s.
It is important that $n-n'$ mixing
cannot destabilize nuclei and thus nuclear
stability limits do not apply in this case.\footnote{
Compare with neutron - antineutron ($n-\tilde{n}$) oscillation
\cite{nnbar}: the direct limit from the neutron experiments in free flight
gives $\tau_{n\tilde{n}} > 10^8$ s \cite{Baldo}
while the nuclear stability tests yield even stronger bounds.
}
From the theoretical side,  $n-n'$ mixing can be induced
from the effective six-fermion operators like $(1/M)^5 (udd)(u'd'd')$
between the ordinary quarks $u,d$ and their mirror twins $u',d'$,
with $\dm \sim (10\, {\rm TeV}/M)^5 \times 10^{-15}$ eV,
$M$ being the relevant cutoff scale.
While the underlying TeV scale physics can be accessible at
the LHC,  the $n-n'$ oscillation itself can have interesting and
testable astrophysical implications,
e.g. for the propagation of ultra-high energy cosmic rays
\cite{BB-nn} or for the neutrons from solar flares \cite{Nasri}.
Moreover, it can be experimentally tested
with slow neutron facilities \cite{BB-nn} and
in particular with the ultra-cold neutron (UCN) storage chambers
(for relevant discussions, see also \cite{Pokot,Kerbikov}).

As far as mirror neutrons are sterile with respect to ordinary
interactions, $n\to n'$ transition can only
manifest as anomalous disappearance of the neutrons,
in addition to the decay, absorption and other regular channels 
of their losses. However, in contrast to the latter, the neutron losses 
due to $n-n'$ transition depend on the magnetic field.

As far as the ordinary and mirror neutrons have equal masses
and decay widths and also their gravitational potentials are universal,
the non-relativistic Hamiltonian describing  the $n-n'$ oscillation in the
vacuum has the form
\be{Hn}
H_I = \mat{ \mu \vec{B}\! \cdot\! \bsig }{\dm} {\dm}
{ \mu  \vec{B}'\! \cdot\! \bsig }   \;,
\ee
where  $\mu = -6 \cdot 10^{-12}$ eV/G
is the magnetic moment of the neutron,
$\vec{B}$ and $\vec{B}'$ respectively are
the ordinary and mirror magnetic fields, and
$\vec{\sigma}=(\sigma_x,\sigma_y,\sigma_z)$
are the Pauli matrices.

It was naively assumed in \cite{BB-nn}  that
there is no significant mirror magnetic field
at the Earth, $B'=0$.
Then non-zero $B$ introduces
the level splitting in the Hamiltonian (\ref{Hn})  that
corresponds to the energy (frequency) scales
\be{om-magn}
2 \om = | \mu B |  = 6 \cdot 10^{-12}
\left(\frac{B}{1 \, {\rm G} }\right)  {\rm eV} =
9 \cdot 10^3 \left(\frac{B }{1 \, {\rm G} }\right) {\rm s}^{-1}  .
\ee
Thus, for the angle of $n-n'$ mixing we have
$| \tan2\theta | = \dm/\om$  and the oscillation frequency is
$\Om = \sqrt{\om^2 + \dm^2}$.
The probability of $n\to n'$ transition after the flight time $t$ reads
\be{PB-t}
P_B(t) =
\sin^2\!2\theta  \sin^2 (\Om t)  =
\frac{\dm^2}{\om^2 + \dm^2}
\sin^2 \left( \sqrt{\om^2 + \dm^2} \, t \right)    .
\ee
Clearly, $P_B$ depends on the magnetic field strength but
does not depend on its orientation. Even if $\dm t \ll 1$, 
the time-oscillating term can be averaged
in strong magnetic field, when $\omega t \gg 1 $,
and the mean oscillation probability reads $P_B = \dm^2/2(\om^2$.

On the other hand, once $B'=0$, the Hamiltonian (\ref{Hn})
becomes degenerate in the limit $B=0$.
Then $n-n'$ mixing is maximal ($\theta_0=45^\circ$),
the oscillation frequency is $\Om_0 = \dm$
and the oscillation probability becomes  $P_0(t) = \sin^2(\dm t )$.
If the neutron free flight is long enough, $t \gg \tau_{nn'}$,
the $n-n'$ transition probability averaged over many oscillations
becomes $1/2$.

In the real experiments, the mean free flight time
of the neutrons is limited by technical reasons. So far up to
$t\sim 0.1$ s can be achieved for the cold neutrons propagated
at distances $\sim 100$ m \cite{Baldo} or in the UCN  traps
 of the dimensions $\sim 1$ m as in the experiments
 \cite{Ban,Serebrov,Serebrov-New}.\footnote{
 The DUSEL project \cite{DUSEL}
 can significantly increase the neutron free flight time.}
For $t \ll \tau_{nn'}$,  the oscillation probability in zero
magnetic field is $P_0(t) = (\dm t)^2\ll 1$.
In the weak magnetic field, with $\om \ll t^{-1}$,
the oscillation probability remains essentially the same,
$P_B(t) =  P_0(t) = (\dm t)^2$.
Namely, for $t \sim 0.1$ s  it would be the case for
$B< 10^{-3}$ G.\footnote{The experimental limit \cite{Baldo}
on the deficit of cold neutrons propagating in a weak magnetic field
$B \sim 10^{-4}$ G with a mean time $t\sim 0.1$ s
implies $P_0(t) = (\dm t)^2 < 10^{-2}$, and thus imposes
the bound $\tau_{nn'} > 1$ s or so \cite{BB-nn}.}
However, if the magnetic field is enough strong,
$\om > t^{-1} $, it should suppress the $n-n'$ transition:
$P_B(t) = (\dm/\om)^2 \sin^2(\om t) \ll (\dm t)^2$.
Therefore, experiments that compare the
neutron losses in the conditions of zero
({\it i.e. weak}, $\om t \ll 1$)
and non-zero ({\it i.e. strong}, $\om t \gg 1$)
magnetic fields in otherwise similar conditions
can directly trace the difference $P_B - P_0$
since the uncertainties related to the neutron decay
and other regular channels can be canceled out.

Three experiments \cite{Ban,Serebrov,Serebrov-New} of this kind
were performed during the last years at the Institute of Laue-Langevin
(ILL), Grenoble. The UCN losses were compared
for different configurations of the magnetic fields and
the limits 
\beqn{limits}
 & \tau_{nn'} > 103 \; {\rm s}  &  \quad (95 \% \; {\rm CL} )  \quad  
\nonumber \\
 & \tau_{nn'} > 414 \;{\rm s}  &  \quad (90 \% \; {\rm CL}) \quad  
\nonumber \\
& \tau_{nn'} > 403 \;{\rm s}   & \quad  (90 \% \; {\rm CL} )  \quad  
\eeqn
were reported in refs. \cite{Ban}, \cite{Serebrov} and \cite{Serebrov-New} 
respectively, 
implying an upper bound  $\dm < 2\cdot 10^{-18}$ eV or so.
However, in these experiments  the data
were analyzed taking $B'=0$
and hence assuming that
(a) the probability of $n-n '$ oscillation $P_B$
does not depend on the magnetic field direction;
(b) magnetic field can only suppress the oscillation,
$P_B < P_0$,
so that the UCN  counts should be larger when the
magnetic field is {\it on} than when it is {\it off}.

In the present paper we study the more general case
when the mirror magnetic field $B'$ is non-zero.
This makes the situation very different.
Indeed, if at the Earth $B'\neq 0$ by some virtue,
it cannot be screened in the experiments
and contributes the mirror neutron energy levels
in the Hamiltonian (\ref{Hn}) as $2\om' = |\mu B'|$.
Hence, the $n-n'$ oscillation probability in the limit $ B=0 $,
if $\omega' t \gg1$, can be averaged in time
and it becomes $P_0 = \frac12 (\dm/\om')^2$,
instead of $P_0 = (\dm t)^2$.
Then the non-zero $\vec{B}$
 can either {\it suppress} or {\it resonantly enhance} the $n-n'$ oscillation,
 depending on its strength as well as  on its orientation
 with respect to $\vec{B}'$.
Therefore experimental data on $n-n'$ oscillations
should be interpreted more carefully.
In particular, in the presence of mirror field $B' > 10^{-2}$ G or so,
the experiments \cite{Ban,Serebrov,Serebrov-New} cannot
impose the limits (\ref{limits}),
and as we see below, the $n-n'$ oscillation time $\tau_{nn'}$
can easily be order 1 s or even smaller.

The paper is organized as follows. In section 2 we study the
Hamiltonian (\ref{Hn})  in the general case when both ordinary and
mirror magnetic fields  $\vec{B}$ and $\vec{B}'$ are non-zero and
arbitrarily oriented. We derive the exact formula for
the $n-n'$ oscillation probability which essentially
depends on two parameters: relative strength of the fields,
$\eta = B/B'$, and the angle $\beta$ between the vectors
$\vec{B}$ and $\vec{B}'$.   In section 3 we
discuss the implications of our results for the UCN storage
experiments and re-visit the experimental data reported in refs.
\cite{Ban,Serebrov,Serebrov-New}.
Namely, in the experiments \cite{Ban,Serebrov-New}
the UCN losses were measured for the vertical directions of the
magnetic field, up ($B_\up$) and down ($B_\down$),
respectively at $B=0.06$ G \cite{Ban}
and $B=0.2$ G \cite{Serebrov-New}.
Interestingly, these experimental data indicate a deviation from zero
for the up-down asymmetry of the neutron losses,
at about $3\sigma$ level.
If these asymmetries are real, they may
indicate the presence at the Earth of a mirror magnetic field $B'$
in the range around   $0.1 - 1$ G,
with a significant vertical component.
The mechanisms that could generate
mirror magnetic fields on the Earth, in solar system or in the Galaxy
and their implications for the time variation of the signal
are discussed in section 4. In addition, the implications of the possible
matter effects or any other spin-independent effects that may lift the
degeneracy between the $n$ and $n'$ states are also studied and
the relevant formulas for the oscillation probabilities are given.
 Finally, we briefly discuss the possible strategies  to search
for the disappearance of the unpolarized or  polarized neutrons 
due to the $n \to n'$ oscillation and for the neutron regeneration 
$n \to n' \to n$, as well as for 
measurements of the neutron precession frequencies
as a function of the magnetic field strength and its orientation.

\section{$n-n'$ oscillation in the background of ordinary 
and mirror magnetic fields}

Let us study the free neutron - mirror neutron oscillation
in general case, when ordinary and mirror magnetic fields,
$\vec{B}$ and $\vec{B}'$,
are both non-zero and have arbitrary orientations.
Denoting $\mu \vect{B}=2\vect{\om}$ and
$\mu \vec{B}' = 2\vect{\om}'$,
the Hamiltonian (\ref{Hn}) can be rewritten as
\be{HI}
H_I = \mat{ 2\, \vect{\om}\,  \bsig }{\dm} {\dm}
{ 2 \, \vect{\om}'  \bsig } =
\mat{  (\vect{b} -\vect{a})\, \bsig }{\dm} {\dm}
{ (\vect{b} +\vect{a}) \, \bsig } ,
\ee
where we introduce the combinations
$\vect{b} =\vect{\om}' + \vect{\om}$
and   $\vect{a}= \vect{\om}' - \vect{\om}$.
We can choose the basis of wavefunctions
$(\psi_+,\psi_-,\psi'_+, \psi'_-)$
corresponding to $n$ and $n'$ states with the spins parallel ($+$)
or antiparallel ($-$) to the direction of vector $\vect{b}$
taken as $z$-axis: $\vect{b} = (0,0,b)$,
while the vector $\vect{a}$ is  taken in the $xz$ plane:
$\vect{a}=(a_x,0,a_z)$,  so that
$a_z b = \vect{a}\vect{b} =  \om^{\prime2} -  \om^2$,
$a_x b = |\vect{a}\times \vect{b}| =
2|\vect{\om} \times \vect{\om}'| $ and
$b=|\vect{\om}'+\vect{\om}| =
[\om^2 + \om^{\prime 2} + 2\vect{\om}\vect{\om}']^{1/2}$.
Hence, in this basis  $\vect{b}\, \bsig =  b\, \sigma_z$,
$\vect{a}\, \bsig = a_x \sigma_x + a_z \sigma_z$,
and  the Hamiltonian has the explicit form of the $4\times4$ matrix
\be{mat44}
H_I = \PM{ b -a_z & -a_x & \dm & 0 \\
-a_x & -b+a_z  & 0 & \dm \\
\dm & 0 & b+a_z & a_x \\
0 & \dm & a_x & -b-a_z }   .
\ee
It can be exactly diagonalized by the unitary transformation
\be{matfin}
H_{I} \to S^\dagger H_I S =
\PM{ 2\tilde{\om}  & 0 & 0 & 0 \\
0 & - 2 \tilde{\om}   & 0 & 0 \\
0 & 0 & 2 \tilde{\om}'   & 0 \\
0 & 0 & 0 & -2 \tilde{\om}' } 
\ee
using the mixing matrix of the form
\beqn{matS}
 S = 
 \PM{ \cos\theta & 0 & \sin\theta & 0 \\
0 & \cos\theta & 0 & -\sin\theta \\
-\sin\theta & 0 & \cos\theta & 0 \\
0 & \sin\theta & 0 & \cos\theta } 
\times \quad \quad \quad 
 \nonumber \\ 
\PM{ \cos\phi & \sin\phi  & 0 & 0\\
-\sin\phi & \cos\phi & 0 & 0 \\
0 & 0 & \cos\phi' & -\sin\phi' \\
0 & 0 & -\sin\phi' & \cos\phi' }\,
\eeqn
with the angles defined as follows:
\beqn{angles}
 \tan 2\theta = \frac{\dm}{a_z}\, ,  
 \quad \quad 
 \tan2\phi =  \frac{a_x}{b-\sqrt{a_z^2+\dm^2}}\, , 
\nonumber \\ 
 \tan2\phi' =  \frac{a_x}{b+ \sqrt{a_z^2+\dm^2}}\,  . 
\eeqn
As for the eigenvalues, we obtain:
\beqn{tilde-om}
&&  2\tilde{\om} = \frac{b-\sqrt{a_z^2+\dm^2}}{\cos 2\phi }  \quad \quad \quad
\nonumber \\ 
&&  = \sqrt{ 2(\om^2 + \om^{\prime 2}) + 2(\om^2 - \om^{\prime 2})
\sqrt{1+ \tan^2 2\theta} + \dm^2 }  , \quad \quad
\nonumber \\
&&  2\tilde{\om}' = \frac{b+\sqrt{a_z^2+\dm^2}}{\cos 2\phi' }  \quad \quad \quad
\nonumber \\ 
&&  = \sqrt{ 2(\om^2 + \om^{\prime 2}) - 2(\om^2 - \om^{\prime 2})
\sqrt{1+ \tan^2 2\theta} + \dm^2 }  . \quad \quad
\eeqn
Therefore, the probability of $n-n'$ transition after the flight time
$t$  reads:\footnote{Here the partial probabilities
of the transitions $n\to n'_+$ and $n\to n'_-$  are summed up,
and so $P_{\vec{B}}(t)$ does not depend on the initial neutron polarization.
}
\beqn{P-t}
P_{\vec{B}}(t) =
\sin^2\! 2\theta  \big[
\cos^2(\phi-\phi')\sin^2(\Om^- t ) + \quad 
\nonumber  \\
 \sin^2(\phi-\phi')\sin^2(\Om^+ t )  \big]  ,
\eeqn
where two characteristic frequencies are respectively
$\Om^\pm =  | \tilde{\om}'  \pm \, \tilde{\om}  |$.
The magnitude of $n-n'$ oscillation is essentially determined
by the angle $\theta$
while the angles $\phi,\phi'$ merely describe the spin precessions:
in fact, they do not enter in the
averaged oscillation  probability  $\ov{P}_{\vec{B}} = \frac12 \sin^2\!2\theta$.
From (\ref{angles}) we obtain:
\be{tan22}
\tan^2\! 2\theta =
\frac{\dm^2 (\vect{\om} + \vect{\om}')^2}
{ ({\om}^2 - {\om}^{\prime2})^2 }
= \frac{ 1+\eta^2 + 2\eta \cos\beta}
{ (1-\eta^2)^2  } \,\tan^2\!2\theta_0  , \quad 
\ee
where $\eta = \om/\omp=B/B'$,
$\beta$ is the angle between the vectors $\vec{B}$ and $\vec{B}'$
and   $\tan 2\theta_0 =\dm/\om'$ stands for
$n-n'$ mixing angle in the limit $B=0$.
In this limit the oscillation frequency is
$\Om_0= ( \omp^2 + \dm^2 )$ and the $n-n'$
oscillation probability becomes
\be{P-t0A}
P_{0}(t) = \sin^2\!2\theta_0 \sin^2 (\Om_0 t) = 
\frac{\dm^2}{\omp^2 + \dm^2}
\sin^2 \! \left( \sqrt{\omp^2 + \dm^2} \, t \right) .
\ee
Rewriting eq. (\ref{tan22}) as
\be{sin22}
\sin^2\! 2\theta = \frac{\sin^2\!2\theta_0 }
{ \sin^2\!2\theta_0  \,+\, \frac{(1-\eta^2)^2  }
{ 1+\eta^2 + 2\eta \cos\beta } \, \cos^2\!2\theta_0 } \, ,
\ee
we see that $\sin^2\!2\theta$ can be {\it smaller}
or {\it larger} than $\sin^2\!2\theta_0$,
depending on the values $\eta$ and $\cos\beta$.
If $\eta > 2$, we have $\sin^2\!2\theta < \sin^2\!2\theta_0$ for
any value of $\cos\beta$.
However, we get $\sin^2\!2\theta > \sin^2\!2\theta_0$
if $\eta < 2$ and $2\cos\beta > \eta^3 - 3\eta$.
(If $\eta < \sqrt 3$ the latter condition can be satisfied
also for  {\it negative} $\cos\beta$, i.e. $\beta > \pi/2$.)
The angle of $n-n'$ mixing can be resonantly amplified
if $\eta$ is enough close to 1.
Namely, if $\eta = 1$ ($B=B'$ exactly), we have maximal mixing,
$\sin^2\!2\theta = 1$, for any angle $\beta\neq \pi$.\footnote{
This is a rather interesting feature of the system described by the
Hamiltonian (\ref{HI}):  the resonance (level-crossing)
condition does not require the exact coincidence
of the vectors $\vect{\om}$ and $\vect{\om}'$;
it is sufficient that their modules are equal, $\om=\omp$,
while their directions can be different.
}
However, the width of the resonance depends on $\beta$.
Namely, inspection of eq. (\ref{sin22}) gives that for arbitrary
$\beta$, the resonance half-width at half-maximum is
$\gamma_{\rm res} = |\tan 2\theta_0 \cos(\beta/2) |$.
It becomes maximal, $\gamma = |\tan 2\theta_0|$,
when the vectors $\vec{B}$ and $\vec{B}'$
are parallel ($\beta=0$), gets smaller for non-zero $\beta$
and vanishes for $\vec{B}$ and $\vec{B}'$ being antiparallel
($\beta=\pi$).
Hence, for these limiting cases we have
\beqn{Ppm}
\sin^2\!2\theta_{(\beta=0)} =
\frac{\tan^2\!2\theta_0}{\tan^2\!2\theta_0+ (1 - \eta)^2} ,
\nonumber \\ 
\sin^2\!2\theta_{(\beta=\pi)}= \frac{\tan^2\!2\theta_0}
{\tan^2\!2\theta_0+ (1 + \eta)^2} ,
\eeqn
while for the case of orthogonal $\vec{B}$ and $\vec{B}'$
($\beta =\pi/2$) we have
\be{Pp2}
\sin^2\!2\theta_{(\beta=\frac{\pi}{2})} =
\frac{\tan^2\!2\theta_0}{\tan^2\!2\theta_0+
(1 - \eta^2)^2\!/(1+\eta^2)}
\,
\ee
which is larger than $\sin^2\!2\theta_0$ if  $\eta < \sqrt3$.

The mirror magnetic field $\vec{B}'$, if it exists at the Earth
by some circumstances,  cannot be screened in the experiments.
Then the probability of $n-n'$ transition should depend
on the magnitude and orientation of the ordinary magnetic
field $\vec{B}$,  provided that the neutron free flight time $t$
is enough large.\footnote{
For very small $t$ the oscillation probabilities would not depend
on magnetic fields. Namely,  for $\Om^+ t \ll 1$
eq. (\ref{P-t}) reduces to  $P(t) \approx (\dm\,t)^2$.
}
In particular, if  $\om' t \gg 1$,
the oscillating term in the transition probability
$P_0(t)$ (\ref{P-t0A}) can be averaged in time, and we obtain
\be{P-0av}
P_0 =  \frac12\, \sin^2\! 2\theta_0 =
\frac{\dm^2}{2\om^{\prime2}} \, .
\ee
E.g., for $t \sim 0.1$ s we have $\om' t > \pi$
if  $2\om' > 5\times 10^{-14}$~eV, or $B' > 7 $ mG.
In addition,  $P_0 \ll 1$ implies that $\om' \gg \dm$.
In particular, the bound on neutron losses in the experiment
\cite{Baldo}
yields the limit $P_0 < 10^{-2}$ or so \cite{BB-nn}.
On the other hand, assuming that $B$ is far enough from the
resonant value $B=B'$, so that $\theta \ll 1$ and
$(1-\eta)^2 \gg P_0$,
for the average oscillation probability (\ref{P-t})
in the magnetic field $\vec{B}\neq 0$ we get
\beqn{PB}
&& P_{\vec{B}} = \frac12 \sin^2\!2\theta =
P_0 \,\frac{1+\eta^2 + 2\eta \cos\!\beta}{(1-\eta^2)^2 }
= P_{B_\bot}\! + D_B\! \cos\!\beta , 
\nonumber  \\
 && P_{B_\bot} =
P_0 \, \frac{1+\eta^2}{(1-\eta^2)^2 } \,  ,
\quad \quad D_B =
P_0 \, \frac{2 \eta }{(1-\eta^2)^2 } \, , 
\eeqn
where $P_{B_\bot}$ corresponds to the case when
$\vec{B}$ and $\vec{B}'$ are orthogonal, i.e. $\cos\beta = 0$.
Hence, the experimentally measurable difference of the
probabilities
\beqn{diff}
P_{\vec{B}} - P_0
= P_{B_\bot}\! - P_0 + D_B\! \cos\!\beta  \quad \quad 
\nonumber \\ 
= P_0 \, \frac{\eta^2(3-\eta^2) + 2\eta \cos\!\beta}{(1-\eta^2)^2 }
\eeqn
depends on the orientation of the magnetic field $\vec{B}$
(angle $\beta$).
Changing the magnetic field direction to the opposite,
$\vec{B} \to - \vec{B}$, while $\vec{B}'$ remains fixed,
i.e. $\beta \to \pi - \beta$,
one sees that $P_{-\vec{B}} \neq P_{\vec{B}}$
unless $\vec{B}$ and $\vec{B}'$ are orthogonal. 
However, the average between $P_{\vec{B}}$ and $P_{-\vec{B}}$
does not depend on $\beta$ 
while their difference is proportional to $\cos\beta$:
 \be{D}
\frac{ P_{\vec{B}} + P_{-\vec{B}} }{2} =
 P_{B_\bot}  , 
\quad \quad 
P_{\vec{B}}  - P_{-\vec{B}} = 2D_B \cos\!\beta \, ,
\ee
So, it is convenient  to measure experimentally the latter difference
as well as
\be{Delta}
\Delta_B  = P_{B_\bot}\! - P_0 =
P_0  \, \frac{\eta^2(3 - \eta^2)}{(1-\eta^2)^2 } \, .
\ee
The sign of $\Delta_B$ depends on the strength of the magnetic 
field $B$. Namely,
it is {\it positive} for weaker fields, when $B < \sqrt3 B'$ and
becomes {\it negative} for stronger fields, $B > \sqrt3 B'$.
For $B \ll B'$ the stronger effect should be observed by measuring
$D_B$ which is nearly linear in $\eta$ while $\Delta_B$ is quadratic.
For the ratio of two effects we get
\be{DD}
\frac{\Delta_B}{| D_B |} =
\frac{1}{2}\, \eta(3- \eta^2).
\ee
This ratio reaches  its maximal value $1$
at $\eta =1$;  vanishes at  $\eta = \sqrt 3$,
turns to the value $-1$ at $\eta = 2$
and rapidly decreases further with increasing $\eta$.
Hence, for $B > 2B'$ the dominant effect should be
observed by measuring $\Delta_B$.

The formulas above assume that $n-n'$ oscillation is far from the
resonance regime. At the resonance, when
$\om'=\om \gg \dm$,  $\sin^22\theta=1$ and
for the oscillation probability we get
\beqn{Pt-res}
P_{\vec{B}}(t) =
\frac{4\om^2-\dm^2 \cos^2\frac{\beta}{2}}
{4\om^2 - \dm^2 \cos\beta } \,
\sin^2\left(t \, \dm \cos \frac{\beta}{2} \right) + \quad 
\nonumber \\ 
\frac{\dm^2 \sin^2\frac{\beta}{2}}{4 \om^2 - \dm^2 \cos\beta} \,
\sin^2\left(t \, \sqrt{4\om^2 + \dm^2 \sin^2\frac{\beta}{2}} \right )  .
\eeqn
Hence, if $\om' t \gg 1$ but
$\dm t \ll 1$, the dominant contribution comes from
the first term which cannot be averaged in time
and so $P_{\vec{B}}(t) \approx \cos^2(\beta/2) (\dm t)^2$
and $P_{-\vec{B}}(t) \approx \sin^2(\beta/2) (\dm t)^2$,
 Therefore,  we get
$D_B = P_{B_\bot} = (t/\tau_{nn'})^2 \gg 2P_0 = (1/\om' \tau_{nn'})^2$,
where $\tau_{nn'}=\dm^{-1}$, and hence at the resonance 
$\Delta_B/|D_B| = 1$,
in agreement with (\ref{DD}).
For example, for $\tau_{nn'}\sim 10$ s and $t=0.1$ s,
we would have $\Delta_B \sim 10^{-4}$.

Concluding this section, let us remark that $n-n'$ mixing
in the background of the mirror magnetic field
should affect also the neutron spin precession.
For the Hamiltonian (\ref{mat44})
the probability of the neutron polarization change
(transition from $\psi_+$ to $\psi_-$ state) reads
\beqn{spin}
P_{\rm pr}(t) & = &
\cos^4\! \theta \sin^2\! 2\phi \sin^2(2\tilde{\om} t) + \quad \quad 
\nonumber \\
&&  \frac12 \sin^2\! 2 \theta \sin 2\phi \sin 2\phi'
\sin(\tilde{\om} t) \sin (\tilde{\om}' t) + \quad 
\nonumber \\ 
 && \sin^4\! \theta \sin^2\! 2\phi' \sin^2(2\tilde{\om}' t) \
\eeqn
In the limit $\theta =0$ (no $n-n'$ mixing),
we have $P_{\rm pr}(t) =\sin^2 2\phi \sin^2(2\om t)$,
the Larmor precession
with normal frequency $2\om=|\mu B|$.
However, for non-zero $\theta$ the precession frequency is affected.
(In particular, the neutron gets a non-zero spin precession
even if the magnetic field is exactly zero, $B=0$: this is an evident
consequence of the mixing with the neutron state while the spin
of the latter precesses with respect the mirror magnetic field
$\vec{B}'$.)
Namely, for $\theta \ll 1$ eq. (\ref{tilde-om}) leads to
\be{t-om}
\tilde{\om}_{\vec{B}} =
{\om} \left[
1 + \frac{P_0 + (\eta^2-1)P_{\vec{B}}}{4\eta^2} \right]  =
\om \, + \,
\frac{\eta + \cos\!\beta}{2 (\eta^2-1)} P_0 \om' .
\ee
From the experimental point of view,
the effect can be detected by comparing the results
of the neutron magnetic moment measurements at
different values and/or directions of the magnetic field.
In particular, one can measure the difference
between the precession frequencies
$\delta \tilde\om_{\vec{B}} = 
\tilde{\om}_\vec{B} - \tilde{\om}_{-\vec{B}}$
for the magnetic fields of the opposite 
direction $\vec{B}$ and $-\vec{B}$.
In doing so, we expect
\be{Delta-om}
\frac{\delta \tilde{\om}_{\vec{B}}}{\om}  =
\frac{\eta^2 - 1}{4\eta^2}  \left[P_{\vec{B}} - P_{-\vec{B}} \right] =
 \frac{\eta^2 - 1}{2\eta^2} D_B \!\cos\!\beta =
  \frac{P_0\cos\!\beta}{ \eta (\eta^2 - 1)} \, .
\ee
These shifts of the precession frequency can be measured
in the experiments similar to the ones designed for a search
of the neutron EDM in which the relative orientation
of the applied electric and magnetic fields are tested.
However, in these experiments usually the electric field
direction is altered  while the magnetic field is kept fixed.
In our case no electric field is needed and only
the  magnetic field should be applied, altering its direction
from $\vec{B}$ to $-\vec{B}$.

\section{$n-n'$ oscillation in the UCN storage chambers }

In the UCN storage chamber $n-n'$ oscillations are restricted
by the free flight time $t_f$ between the neutron collisions on the
walls.
During the free flight the initial neutron state gets
a small admixture of mirror neutron state
and thus per each wall collision
it can escape from the trap with a mean probability $P$
which is equal to the $n-n'$ oscillation probability
averaged over the neutron distribution.
Therefore, given that  the initial amount of the neutrons
$N_{\rm in}$ is fixed, the amount of the neutrons that
remain in the trap after the storage time $t_s$ should be
$N(t_s) = N_{\rm in}
\exp\big[-(\Gamma + R_s + P \nu) t_s\big]$,
where $\nu = 1/t_f$ is a mean collision frequency,
$\Gamma=\tau_{\rm dec}^{-1}$ is the neutron decay width
and $R_s$ accounts for the regular UCN losses due to absorption
and upscattering during the wall collisions or in collisions
with the residual gas in the chamber.
The neutron losses due to $n-n'$ oscillation can be smaller
than the latter effects. However, if the $n-n'$ oscillation
probability depends on the magnetic field,
then by comparing the neutron counts $N_1(t_s)$ and $N_2(t_s)$
measured for two configurations $\vec{B}_1$ and $\vec{B}_2$
with different strength and/or orientation of the magnetic field,
the dependence on $\Gamma$ and $R_s$ cancels out and the
ratio $N_1(t_s)/N_2(t_s) = \exp\big[(P_2-P_1) \nu t_s \big]$
directly traces the difference between the oscillation probabilities
$P_1=P(\vec{B}_1)$ and  $P_2=P(\vec{B}_2)$.

In particular, one can measure the neutron counts $N_{\vec{B}}$
for an applied magnetic field $\vec{B}$ after a holding time $t_s$,
then change the direction of the magnetic field and
measure the neutron counts $N_{-\vec{B}}$ in the same conditions.
The expected directional asymmetry
of the neutron counts can be simply calculated and we get
\beqn{asymmetry}
 A(t_s) =
\frac{N_{\vec{B}}(t_s) - N_{-\vec{B}}(t_s)}
{N_{\vec{B}}(t_s)+N_{-\vec{B}}(t_s)} = 
 \frac {e^{-n_s P_{\vec{B}} } - e^{-n_s P_{-\vec{B}} } }
{ e^{ -n_s P_{\vec{B}} } + e^{-n_s P_{-\vec{B}} } } 
\quad \quad
\nonumber \\ 
= - \tanh(n_s  D_B \cos\!\beta  ) \quad 
\eeqn
where $n_s = \nu t_s$ is the mean amount of wall collisions
per neutron during the time $t_s$, and
$P_{\vec{B}} - P_{-\vec{B}} = 2 D_B \cos\!\beta$ (see eq. (\ref{D})).
On the other hand, one can compare the neutron
counts for zero magnetic field, $N_0$, with
the averaged counts between the opposite directions of the
non-zero magnetic field, $N_B = \frac12 (N_{\vec{B}}+N_{-\vec{B}})$.
Then we have
\beqn{EB}
1 + E(t_s) =
\frac{N_0(t_s)} {N_B(t_s)} = 
\frac{2 e^{- n_s P_0 } }
{ e^{-n_s P_{\vec{B}}  } + e^{ -n_s P_{-\vec{B}}  } }  \quad \quad 
\nonumber \\ 
= \frac {\exp(n_s  \Delta_B ) } { \cosh(n_s D_B \cos\!\beta ) } \quad 
\eeqn
where $\Delta_B = \frac12 (P_{\vec{B}} + P_{-\vec{B}}) - P_{0}$
(see eq. (\ref{Delta})).
For small oscillation probabilities, when
$D_B,\Delta_B\ll n_s^{-1}$,
we have approximately $ A(t_s) = - n_s D_B \cos\beta $ and
$E(t_s) = n_s \Delta_B $,
and hence the ratio $E(t_s)/ | A(t_s) | = \Delta_B/ | D_B |$
should not depend on the holding time $t_s$.

The experiment \cite{Ban} used the UCN chamber where
the mean free flight time between the wall collisions
was $t_f \simeq 0.05$ s and hence
$\nu = t_f^{-1} \simeq 20 \, {\rm s}^{-1}$.
The measurements were performed for the storage times
$t_s^\ast = 50$ s, 100 s and 175 s and the neutron counts
were compared in the conditions of ``zero" ($B_0 < 0.5$ mG)
and "non-zero" ($B = 0.06$ G) magnetic fields.
The direction of the latter
was vertical and altered from up ($B_\up$) to down ($B_\down$)
repeating the configuration sequence:
$B_0,B_\up,B_\down,B_0,B_0, B_\down,B_0,B_\up$.
Another sequence $B_\up,B_\down,B_0$ was also used 
for $t_s^\ast = 50$ s. 
Taking into account that the neutrons can oscillate also during the 
filling and emptying of the chamber, 
the effective holding time was estimated as
$t_s = t_s^\ast + 23 $ s.

Results of the measurements \cite{Ban} are reported in Table 1.
They indicate that the neutron counts in the non-zero magnetic field
depend on the direction of the latter.
We observe that $N_{B\up}$ regularly exceeds $N_{B\down}$
for all 4 cycles of data taking.
As far as under the naive assumption $B'=0$ \cite{BB-nn}
this feature was not expected,
since in this case the probability of $n-n'$ oscillation
is independent of the magnetic field direction,
in ref. \cite{Ban} this effect was neglected,
the counts $N_{B\!\up}$ and $N_{B\!\down}$ were
averaged and compared to $N_0$.
As a result, the first limit in (\ref{limits}) was imposed.
However, in the presence of mirror magnetic field $\vec{B}'$
the difference $N_{B\!\down} - N_{B\!\up}$ must
depend on the angle $\beta_V$ between $\vec{B}'$ and $\vec{B}_\down$.
Thereby, let us analyze  the data of ref. \cite{Ban}
allowing the oscillation probabilities
$P_{B\down}$ and $P_{B\up}$ to be different.
Then, fitting $E(t_s)$ and $A(t_s)$
given in Table \ref{tab:Ban}, 
we obtain within $1\sigma$ error-bars:
\be{Delta-fit}
 \Delta_{[B=0.06 \,{\rm G}]}  = (2.9 \pm 4.4) \times 10^{-7} , \quad 
 \chi^2_{\rm d.o.f.} = 6.9/3    
\ee
and 
\be{D-fit}
  D_{[B=0.06 \,{\rm G}]}\cos\!\beta_V = (6.2 \pm 2.0) \times 10^{-7} ,
\quad   \chi^2_{\rm d.o.f.} = 0.52/3  
\ee
As we see, $D_B$ has  a $3\sigma$ deviation from zero while
there is no pronounced effect for $\Delta_B$.\footnote{
The bad $\chi^2$ for $\Delta_B$, is due to the result for
$E(t_s)$  for  $t_s = 50$ s in Table 1
which in ref. \cite{Ban} was interpreted as $3\sigma$ fluctuation.
Without taking it into account,  one would get
$\Delta_B = (0.1 \pm 3.1) \times 10^{-7}$ with a vanishing  $\chi^2$.
However, in this experiment the acquired data were rather poor and
hence neither the statistical features for $\chi^2$ should be expected
nor must the error-bars in (\ref{Delta-fit}) and
(\ref{D-fit}) be taken very seriously. }

 \begin{table*}
 \begin{center}
\def\arraystretch{1.2}
\begin{tabular}{c||c|c|c|c}
$t_s$
&  73 s  &  73 s $^{\dagger}$ & 123 s & 198 s     \\
\hline
\hline
$N_{B\!\up}(t_s)$
& $44197 \pm 53$  & $44443 \pm 53$
& $28671 \pm 30$  &  $17047 \pm 31$    \\
$N_{B\!\down}(t_s)$
& $44128 \pm 53$  & $44316 \pm 46$
& $28596 \pm 30$  & $16974 \pm 31$    \\
$N_0(t_s)$
& $44317 \pm 40$  & $44363 \pm 53$
& $28635 \pm 21$  & $17015 \pm 22$   \\
\hline
$E(t_s)\times 10^3 $
& $3.50 \pm 1.24 $ & $ -0.37 \pm 1.43 $
& $0.05 \pm 1.04 $  & $ 0.27 \pm 1.83$    \\
$A(t_s) \times 10^3$
& $0.78 \pm 0.85 $ & $ 1.43 \pm 0.79 $
&  $1.31 \pm 0.74 $  &  $2.15 \pm 1.28$  \\
\end{tabular}
\caption{The UCN counts reported in ref. \cite{Ban}
measured in different sequences of the magnetic field
configurations for different times $t_s$.
The symbol $\dagger$ marks the sequence $B_\up,B_\down,B_0$. 
}
 \label{tab:Ban}
\end{center}
\end{table*}

A more recent experiment \cite{Serebrov-New}
has collected much bigger statistics.
The ``up-down" asymmetry  of the neutron counts
$N_{B\!\up}$ and $N_{B\!\down}$ was measured
for a vertical applied magnetic field $B\approx 0.2$ G,
repeating the configuration sequence:\,
$B_\up,B_\down,B_\down,B_\up;B_\down,B_\up,B_\up,B_\down$. \,
Such a sequence of measurements gives an important advantage
since it allows to remove
a linear drift and also an eventual quadratic drift
in the neutron flux, vacuum conditions, etc..
In addition, the neutron flux was monitored during the filling
of the trap. After the UCN holding time  $t_s^\ast =300$ s,
the neutrons were counted using two independent detectors.
 As a consequence, for an effective exposition time estimated
 as $t_s=370$ s, the following result was obtained:
 $A_V(t_s) = (3.8 \pm 1.2) \times 10^{-4}$.
 If this $3\sigma$ deviation is not related to statistical
 fluctuations or some unknown systematic effects, then it
 may point to the $n-n'$ oscillation
in the background of a mirror magnetic field  $\vec{B}'$
with a significant vertical component: $\cos\beta_V \sim 1$.
The control measurements performed in the UCN flow mode in order
to check whether this deviation was induced by the
influence of the current switching on the electronic systems
have shown no systematic effects of such type
at the accuracy level $10^{-4}$.
Then,  using eq. (\ref{asymmetry}) and
the effective amount of wall collisions per neutron
estimated as $n_s = \nu t_s \approx 4\times 10^3$,
this translates to
\be{D-V}
 D_{[B=0.2 \,{\rm G}]} \cos\!\beta_V = (9.5 \pm 3.0) \times 10^{-8} .
\ee
In the same experiment \cite{Serebrov-New} the asymmetry
of the UCN counts between the configurations $\vec{B}_H$ to $-\vec{B}_H$
was measured also for the horizontal magnetic fields ($B=0.2$ G),
directed roughly towards North-East, with the following result:
$A_H(t_s) = (0.3 \pm 5.1) \times 10^{-5}$, which translates as
\be{D-H}
 D_{[B=0.2 \,{\rm G}]} \cos\!\beta_H = (0.1 \pm 1.3) \times 10^{-8} ,
\ee
with $\beta_H$ being the angle between $\vec{B}_H$ and $\vec{B}'$.
At the same time, for the difference between the UCN counts measured
in "small" ($B < 0.012$ G) and "large" ($B = 0.2$ G) magnetic
fields, that according to (\ref{EB})  should not depend essentially
on the magnetic field orientation, the following result was obtained:
\be{Delta-H}
 \Delta_{[B=0.2 \,{\rm G}]} = -(3.5 \pm 2.5) \times 10^{-8} ,
\ee
In the above considerations it was implicitly assumed that
the angles $\beta_V$ and $\beta_H$ are constant in time.
This would occur e.g. if the mirror magnetic field $\vec{B}'$
is related to the Earth,
rotates with the Earth at the same angular velocity,
and so its orientation with respect to the experimental
site does not change in time.\footnote{
The analysis would be more complicated if $\beta_V$ and $\beta_H$
vary with time,
e.g. if the background field $\vec{B}'$ has a fixed direction
in the solar system while the direction of experimental field $\vec{B}$
changes with the Earth rotation, as discussed in Section 4.
In the experiment \cite{Ban} the time interval between the
configurations $B_\up$ and $B_\down$  was typically few hours,
and hence at the corresponding time moments
the angles $\beta_\down$ and $\beta_\up$ could be different.
(In stable experimental conditions the neutron counts
at zero magnetic field, $N_0$, should
anyway remain time independent.)
Therefore, instead of eqs. (\ref{D}) and (\ref{Delta}) we would get
$D_B = 2P_0 \eta \cos\beta\cos\alpha/(1-\eta^2)^2$ and
$\Delta_B = P_0 (3\eta^2 - \eta^4 - 2\eta \sin\beta\sin\alpha)
/(1-\eta^2)^2$, where $\beta = (\beta_\up + \beta_\down)/2$
and  $\alpha = (\beta_\up - \beta_\down)/2$.
Notice that this could imitate {\it negative} $\Delta_B$
for small $\eta$'s while the truly equal time measurement
of $N_{B\down}$ and $N_{B\up}$ should always
give {\it positive} $\Delta_B$ if $\eta < \sqrt3$
(c.f. eq. (\ref{Delta})).
}
Another possibility is that the Earth itself is the origin
of some pseudo-magnetic potential acting on the $n'$ state,
mediated by some light axion-like fields,
as will be discussed in Section 4.

Let us suppose now, that $3\sigma$ deviation (\ref{D-V})
is not just a fluctuation and it indeed points to
$n-n'$ oscillation.
Then we wonder, how large mirror field $B'$ is required
to explain it?
Using the formulas (\ref{D}), (\ref{Delta}) and (\ref{DD}),
we see that the experimental data (\ref{D-V}), (\ref{D-H})
and (\ref{Delta-H})
are compatible with $\cos\beta_V \simeq 1$, $\cos\beta_V \ll 1$,
and $\eta \simeq 1.8$, i.e. $B'\simeq 0.11$ G.
Namely, the latter estimation follows from comparing the
values of $D_B$ and $\Delta_B$ using the relation (\ref{DD});
namely that the negative $\Delta_B$ implies $\eta > \sqrt3$,
but $| \Delta_B/D_B | < 1$ tells that $\eta < 2$.
(Treating the errors less conservatively, one can
consider a larger interval $B= (0.08 - 0.15)$ G.)
Error bars leave a margin also for positive
$\Delta_B$, however with $| \Delta_B/D_B | < 0.3$ or so.
Thus we have another branch of solution, with
$\eta < 0.2$, or $B' > 1$ G.

How fast $n-n'$ oscillations can be in the absence of
the magnetic fields or at the resonance, i.e. how small
 oscillation times $\tau_{nn'}=\dm^{-1}$ are allowed?
From eqs. (\ref{D}) and (\ref{P-0av}) we obtain
$P_0 = D_B(1-\eta^2)^2/(2\eta)$ and
$\dm =  \om \sqrt{D_B} \cdot |1-\eta^2|/\eta^{3/2}$;
$B=0.2$ G means $\om = 6 \times 10^{-13}$ eV.

For example, let us take a central value
$D_{[B=0.2 \,{\rm G}]} = 9.5 \times 10^{-7}$
and $B'=0.11$ G ($\eta=1.8)$. Then we get
$P_0 = 1.4\times 10^{-7}$,  $\tau_{nn'}=\dm^{-1} = 3.8$ s,
and  also
$D_{[B=0.06 \,{\rm G}]}  = 3 \times 10^{-7}$, which
curiously is compatible with the fit (\ref{D-fit})
following from the experiment \cite{Ban}.
Along the same lines for e.g. $B'=0.15$ GeV we get
$P_0 = 2.1\times 10^{-8}$ and $\tau_{nn'}=5.5$ s.
As for the small $\eta$ branch,  $B'>1$ GeV implies
$P_0 > 2 \times 10^{-7}$ and $\tau_{nn'}< 0.3$ s.

As for another $3\sigma$ deviation (\ref{D-fit}) indicated
by the data of ref. \cite{Ban}, probably it should not be taken
very seriously. Let us discuss, nevertheless, what size of
 mirror magnetic fields it requires. Then the  lower bound
can be settled from the value of $\Delta_B$ (\ref{Delta-fit}).
As one can see from eq. (\ref{Delta}),
$\Delta_B$ could be positive if $B'$
is large enough, namely if $\eta = B/B' < \sqrt3$.
On the other hand, according to eq. (\ref{DD}), for $\eta > 2$
one would expect $\Delta_B$ to be 
negative and larger than $D_B$,
which seems incompatible with the indications
(\ref{Delta-fit}) and (\ref{D-fit}).
Thus, one can set a conservative lower bound
$B' > 0.03 $ G or so. This in turn implies that for the neutron free
flight time being $t_f \approx 0.05$ s in the experiment \cite{Ban},
we have $\om' t_f > 13$ and hence the approximation
of the time-averaged oscillation probabilities is valid.
The upper limit on $B'$ can be imposed from the following
consideration.  As follows from eq. (\ref{D}), for $\eta \ll 1$
we get
$P_0 \approx D_B(2\eta \cos\beta)^{-1} \geq D_B  (B'/0.12 \, {\rm G})$.
Then, assuming $D_B > 4 \times 10^{-7}$,
$B' > 3$ G would imply $P_0 > 10^{-5}$.
On the other hand, for $B' > 3$ G the oscillation probability
$P_{\vec{B}}$ in  the Earth magnetic field
($B\simeq 0.5$ G in Grenoble),
would be in fact larger than  $P_0$ and thus
larger than $10^{-5}$, which is excluded by the upper bound
on the UCN losses in the Earth magnetic field \cite{Sereb}.
Therefore, rather conservatively, one can impose an upper bound
$B' < 3$ G, which by strict analysis can be improved
by a factor of 2 or so.

The experiment \cite{Serebrov} used the horizontally directed
magnetic field $\vec{B}_H$ with $B=0.02$ G, without altering its direction.
Unfortunately, since in this experiment the magnetic field direction was not
altered, the values of $D_B$ and $\Delta_B$ cannot be evaluated.
Its result reads $P_{\vec{B}_H}-P_0 = \Delta_B + D_B \cos\beta_H =
-(1.7 \pm 3.6) \times 10^{-8}$.
The angle $\beta_H$ between $\vec{B}'$ and $\vec{B}_H$
can vary between $\pi/2 \pm \beta_V$,
$\beta_V$ being the angle between $\vec{B}'$ and $\vec{B}_\down$
relevant for the vertical measurements.
Therefore, in spite of much larger statistics, its data
cannot be used directly for our analysis.
Nevertheless, with a careful study, they could provide some additional
information about the orientation of the mirror magnetic field
$\vec{B}'$, more significantly for small $B'$ region.
For example, for $\vec{B}'$ directed nearly
vertically ($\beta< 10^\circ$)
these data would exclude $B'$  less than $0.05$ G or so.

\section{Mirror matter and mirror magnetic fields }

Let us discuss now the origin of the mirror magnetic fields
to find out how strong values are plausible.

Mirror matter can give a substantial fraction
of dark matter in the Galaxy,  or can even entirely represent it.
Thereby, one could naively expect that the mirror magnetic fields
are comparable to ordinary galactic fields $B\sim 10$ $\mu$G
or perhaps even stronger, up to $B' \sim 1$ mG.
Some amplification could occur e.g. in the context of the generation
mechanism \cite{Dolg} for the primordial magnetic field seeds
at the scales of 1 Mpc,
in view of the earlier recombination of mirror matter and its larger
residual ionization  \cite{BCV}.

On the other hand, one cannot exclude that by chance
the solar system is passing presently through a giant molecular
cloud of mirror matter.
It is known that in ordinary molecular clouds the magnetic
fields can be typically order $1-10$ mG and in their dense regions
even up to 100 mG.
Therefore, the mirror magnetic fields up to
100 mG should not look as a surprise.

If the mirror magnetic field has a galactic origin, or it is related to
the mirror molecular clouds, then the vector  $\vec{B}'$
would remain constant in time while the experimental
field $\vec{B}$ rotates together with the Earth, so  that
the angle $\beta$ between $\vec{B}'$ and $\vec{B}$ becomes
a periodic function of time, with a period equal to siderial day.
Therefore, in this case the UCN losses should indicate
specific time variations.
In particular,  the ``up-down" asymmetry of
the neutron counts measured for the applied vertical fields $B_\up$ and
$B_\down$ must have the day-night variations unless
$\vec{B}'$  is by chance parallel  to the Earth rotation axis.
Namely,  we would have
\be{beta-V}
\cos\!\beta_V = \cos\!\gamma \sin\!\varphi  + \sin\!\gamma \cos\!\varphi
\cos \left( 2\pi \frac{t-t_0}{t_{\rm sd}} \right)  ,
\ee
where $\gamma$ is an angle that vector $\vec{B}'$ makes with
 the Earth rotation axis, $\varphi$ is a latitude of
 the experimental site ($\approx 45^\circ$ for Grenoble),
 and $t_{\rm sd}=23.9345$ h is a siderial day.
As for the case of horizontal magnetic fields, directed
e.g. to North or to East, we  respectively get
\beqn{beta-H}
&& \cos\!\beta_N = \cos\!\gamma \cos\!\varphi  - \sin\!\gamma \sin\!\varphi
\cos \left( 2\pi \frac{t-t_0}{t_{\rm sd}} \right)  ,
\nonumber \\
&& \cos\!\beta_E =  - \sin\!\gamma
\sin \left( 2\pi \frac{t-t_0}{t_{\rm sd}} \right)  .
\eeqn
The experimental data \cite{Serebrov,Serebrov-New} can be carefully
analyzed in order to see whether there are the traces of such variations
in $P_{\vec{B}} - P_{-\vec{B}} = 2D_B \cos\beta$,\footnote{
As for the experiment \cite{Ban}, all data
were taken during  the daytime and thus the possibility of
day-night variation instead of the constant fit (\ref{D-fit})
 cannot be {\it a priori} excluded.
}
and certainly new
dedicated experiments in which the strength and orientation of
the applied magnetic field can be varied are desirable.
Let us recall that no variations should be expected
for $\Delta_B$ since $P_{\vec{B}} + P_{-\vec{B}}$
 does not depend on the angle $\beta$.

Another possibility is that the mirror matter is captured by
the solar system and correspondingly in the solar neighborhoods
the substantial mirror magnetic field is present.
One can also expect that in this case $\vec{B}'$ is not
homogeneous in the solar system and it varies
around the Earth orbit (both the strength and the direction).
In this case short time measurements should indicate
a day-night oscillations in $P_{\vec{B}} - P_{-\vec{B}}$
with a fixed $P_{\vec{B}} + P_{-\vec{B}}$, while long time
experiments must show certain annual modulations that would
depend on the pattern of the mirror magnetic field lines
that cross the Earth orbit.

According to common sense,
there should be no significant amount of the mirror matter in the Earth.
The gravitational potential of the Earth is not efficient
to capture a large amount of cosmic mirror particles.
However, the situation could change if there are
some stronger interactions between the ordinary and mirror
matters, e.g. due to photon - mirror photon kinetic mixing
as discussed e.g. in refs. \cite{Mitra}. Such interactions
could give also a consistent explanation to the DAMA/Libra
results on dark matter search \cite{Dama}. On the other hand,
if the neutron - mirror neutron mixing is possible, it would look
pretty natural that also neutral mesons of the ordinary
and mirror sectors have a reasonable mixing:
$\pi  - \pi'$, $\eta-\eta'$, $\rho-\rho'$ etc. that would
mediate enough strong ``nuclear" forces that
could efficiently capture the mirror nuclei in the Earth,
with cross-sections up to few pb.
Interestingly, the geophysical constraints on the amount
of the mirror matter within the Earth appear to be rather
flexible, allowing for up to $0.4$ per cent of the Earth mass
constituted by mirror particles \cite{IV}. If so, then the
existence of the mirror magnetic field $B'$ of order $1$ G
or even larger should not be a surprise, if one takes into 
account that the Earth's rotation itself can
give rise to the asymmetric capture of the mirror matter
that can give rise to circular currents, as well as
the possibility of very efficient dynamo mechanism.
In the view of the latter, the mirror magnetic field could
have time variations much faster than the
terrestrial magnetic field:
the latter changes its polarity in every few million years.
Depending on the interaction strength between the ordinary
and mirror particles as well as on the chemical composition
of the latter, the two following situations can be envisaged:
first, when the mirror matter forms a puffy cloud around the
Earth, with a size much larger than the Earth radius,
that can have a differential rotation. In this case 
the measurements of $P_{\vec{B}} - P_{-\vec{B}}$
can exhibit a quasi-periodic pattern originated by
superposition of the Earth rotation with the slower rotation
of the mirror cloud;
and second, when captured mirror particles form a compact body
inside the Earth that rotates together with the latter
with the same angular velocity. In this case no  time variation
of the signal should be expected.

Another possibility is that the Earth itself is the origin
or some pseudo-magnetic field acting on the $n'$ state.
Imagine, for example,
a light axion-like boson $\chi$ that
has a pseudoscalar coupling with mirror neutrons
$i g \chi \bar{n}'\gamma^5 n =
g(\partial_\mu \chi/m) \bar{n}'\gamma^5 \gamma^\mu n $, but also
has scalar couplings with normal matter components.
Such a hybrid boson would mediate the
long range Yukawa type ``fifth forces"
violating the weak equivalence principle,
and also induce the CP-violating monopole-dipole
interactions discussed in ref. \cite{MW}.
This could occur, if e.g. the mirror axion
having the Yukawa interactions with mirror
baryons \cite{BGG}, is mixed in some way with
a dilaton like scalar coupled to the trace of the
energy-momentum tensor of the normal matter.
(Interestingly, in the mirror gravity scenario
with Lorentz-violation \cite{BCNP}, even the massive
graviton could mix with the axion like scalar).
Then, if $\chi$ is very light, with a Compton length
comparable to the Earth radius, the Earth itself
acts as a source for a spin-dependent static potential
$(g\vect{\nabla}\chi/2m)\, \ov{n}'\vect{\Sigma}\, n'$
where $m$ is the neutron mass, and
$\vec{\Sigma}= {\rm diag}(\vect{\sigma},\vect{\sigma})$ is
the spin matrix, and $\vect{\nabla}\chi$ acts as a vertically
directed pseudo-magnetic field, i.e. $\beta_V = 90^\circ$
and $\beta_H=0$.

The following remark is in order.
In the case of the Earth bounded mirror matter, a non-zero density
of mirror gas can induce a significant spin-independent
contribution in the Hamiltonian of $n'$ state.
(as for the ordinary gas, in the UCN chambers it is pumped out
in order not to affect significantly the neutron propagation).
In this case the  effective Hamiltonian describing the $n-n'$ 
oscillations becomes 
 \be{HI-rho}
H_I = \mat{ v + 2\, \vect{\om}\,  \bsig  }{\dm} {\dm}
{v'   + 2 \, \vect{\om}'  \bsig } , 
\ee
where $2\vect{\omega} = \mu \vec{B}$,  $2\vect{\omega}' = \mu \vec{B}'$,  
and $v$ and $v'$ are the matter induced spin-independent potentials
respectively for ordinary and mirror neutrons.\footnote{
More generally, there can be other reasons that may provide 
different spin-independent
potentials between $n$ and $n'$ states, e.g. if the gravitation
forces are not quite universal between the ordinary and mirror matters
\cite{BCNP}. In fact, the Hamiltonian (\ref{HI-rho}) describes
also a situation when the ordinary and mirror neutrons
are not quite degenerate and their masses $m$ and $m'$
have a small splitting.
}
In the basis of the wavefunctions
$(\psi_+,\psi_-,\psi'_+,\psi'_-)$ where $\psi_\pm$
correspond to the neutron states with the spins parallel/antiparallel
to $\vec{B}$, and $\psi'_\pm$ to
the mirror neutron states with the spins parallel/antiparallel
with respect to $\vec{B}'$, the Hamiltonian has the form
\be{mat44-rho}
H_I =
\PM{2(\rho+\om)  & 0 & \dm \cos\frac{\beta}{2} & -\dm \sin\frac{\beta}{2} \\
0 & 2(\rho-\om) & \dm \sin\frac{\beta}{2} & \dm \cos\frac{\beta}{2}\\
\dm \cos\frac{\beta}{2} &  \dm \sin\frac{\beta}{2} &   2\om'  & 0 \\
-\dm \sin\frac{\beta}{2} & \dm \cos\frac{\beta}{2} & 0 & - 2\om' } , 
\ee
where 
$2\rho = v - v'$  and
$\beta$ is the angle between the vectors $\vect{\om}$ and $\vect{\om}'$.
Hence, the neutron states with the $(+)$ and  $(-)$ polarizations
have different mixings with the mirror neutron states.
Assuming that the mixing angles are small ($\ll 1$), for the time-averaged
oscillation probabilities $n_+ \to n'$ and  $n_- \to n'$ respectively
we obtain:\footnote{Here the
sum is taken over the polarizations of the final $n'$ states.
Clearly, in the limit $\rho = 0$ we have
$P_{_+}(\vec{B}) =P_{_-}(\vec{B}) =P(\vec{B})$, the latter given
by eq. (\ref{PB}).
}
\beqn{PB-rho}
P_{_+}(\vec{B}) & = &
 \frac{\dm^2 \cos^2 \frac{\beta}{2}}{2(\om' -\om-\rho)^2} +
\frac{\dm^2 \sin^2 \frac{\beta}{2} }{2(\om'+ \om +\rho )^2} 
\nonumber \\ 
& = &  \frac{\dm^2 \big[1 + (\eta + y)^2 + 2 (\eta + y) \cos\beta \big] }
{2\om'^2 \big[1 - (\eta + y)^2 \big]^2 } \, ,
\nonumber \\
P_{_-}(\vec{B}) & = &
 \frac{\dm^2 \sin^2 \frac{\beta}{2}}{2( \om' + \om - \rho)^2} +
\frac{\dm^2 \cos^2 \frac{\beta}{2} }{2( \om' - \om + \rho)^2} 
\nonumber \\ 
& = & \frac{\dm^2 \big[1 + (\eta - y)^2 + 2 (\eta - y) \cos\beta \big] }
{2\om'^2 \big[1 - (\eta - y)^2 \big]^2 } \, ,
\eeqn
where  $y = \rho/\om'$ and $\eta = \om/\om'$. Obviously, here we assumed that
neither $P_{_+}(\vec{B})$ nor $P_{_-}(\vec{B})$ are at the resonance.
The difference between $P_{_+}(\vec{B})$ and $P_{_-}(\vec{B})$
can be measured in the experiments with the polarized neutrons.
For unpolarized neutrons, as in the case of experiments
\cite{Ban,Serebrov,Serebrov-New}, two probabilities (\ref{PB-rho})
can be averaged. Thus we obtain:
\be{PB-av}
P_\vec{B} = \frac12 \left[ P_{_+}(\vec{B}) + P_{_-}(\vec{B}) \right] =
P_{B_\bot}\! + D_B \cos\beta
\ee
where
\beqn{PB-D}
&& P_{B_\bot} = \frac{P_{\vec{B}} +P_{-\vec{B}}}{2} =
\frac{\dm^2}{4\om^{\prime 2} }
\left[
\frac{1 + \eta_+^2}{\big(1-\eta_+^2\big)^2}  +
\frac{1 + \eta_-^2}{\big(1-\eta_-^2\big)^2}  \right]  ,  \quad 
\nonumber \\
&& D_B = \frac{P_{\vec{B}}-P_{-\vec{B}}}{2\cos\beta} =
\frac{\dm^2}{4\om^{\prime 2} }
\left[
\frac{2\eta_+}{\big(1-\eta_+^2\big)^2}  +
\frac{2\eta_-}{\big(1-\eta_-^2\big)^2}  \right]  \quad 
\eeqn
($\eta_\pm = \eta \pm y$),  
while the averaged oscillation probability in the limit
of zero magnetic field, $B = 0$, is
\beqn{P0-rho}
 P_0 = \frac12 \left( P_0^+ + P_0^-\right) =
\frac12 \left[ \frac{\dm^2}{2(\rho + \om')^2} +
\frac{\dm^2}{2(\rho - \om')^2} \right]  
\nonumber \\ 
= \frac{\dm^2 }{2\om^{\prime 2}} \cdot \frac{(1 + y^2)}{(1 - y^2 )^2}
\quad \quad 
\eeqn
where we assumed that $y^2\neq 1$, i.e. $P_0$ has no resonance
at $\om = 0$. Thus, the values $D_B$ and
\beqn{Delta-rho}
&& \Delta_B = P_{B_\bot}\! - P_0 = 
\nonumber \\ 
&& \frac{\dm^2 }{2\om^{\prime 2}}
\left[ \frac{\eta_+^2 (3-\eta_+^2)}{2\big(1-\eta_+^2\big)^2}  +
\frac{\eta_-^2 (3-\eta_-^2)}{2\big(1-\eta_-^2\big)^2} -
 \frac{y^2(3- y^2)}{(1 - y^2 )^2} \right]  \quad 
\eeqn
where $y = (\eta_+ - \eta_-)/2$, 
can be tested experimentally with magnetic fields of
varying strength and direction.

New interesting features emerge with respect to the case
$\rho = 0$, when the $n-n'$ oscillation probability had only
one resonance at $\om = \om'$.
Now we have two resonance values of $\om$. Namely, if
$0 < y < 1$, $P_{_+}(\vec{B})$ has a resonance at
$\om = \om' - \rho$ while $P_{_-}(\vec{B})$ has a resonance at
$\om = \om' + \rho$.  For $\om'$ and $\rho$ fixed,
the sign of $\Delta_B$ depends on $\om$, while the sign of
$D_B$ depends on $\beta$ but does not depend on $\om$.
On the other hand, for $y > 1$,  $P_{_-}(\vec{B})$ has two
resonances, at $\om = \rho \pm \om'$ while the sign of $D_B$
changes with increasing $\om$. Notice also, that due to the 
difference between $P_{_+}(\vec{B})$ and $P_{_-}(\vec{B})$,
the UCN with $+$ and $-$ polarizations  should disappear with
the different rates and thus surviving neutrons should have
a preferred polarization even if the neutrons initially  were
unpolarized. These effects can be tested experimentally also
if one varies the value of $\rho= (v-v')/2$ by changing the
residual gas pressure in the UCN traps.

\section{Discussion and outlook}

Summarizing, if a reasonably large mirror magnetic field, 
say $B' > 0.01$ G,  
exists on the Earth or its environments, 
it cannot be screened in the terrestrial experiments
and can strongly affect the neutron to mirror neutron
oscillation features. In particular, the oscillation
probability becomes dependent on the strength
and the direction of the applied magnetic field $B$.
Therefore, the experimental data \cite{Ban,Serebrov,Serebrov-New}
on the $n-n'$ oscillation should be analyzed with more care.
In particular, $\tau_{nn'}$ cannot be anymore restricted by the
limits of about 400 s (\ref{limits}), and in fact it easily
could be of order $1$ s or even smaller.

The issue of a fast $n-n'$ oscillation can have interesting links.
Namely, the questions whether the $n-n'$ oscillation is related to
anomalous neutron losses observed for different material surfaces 
\cite{Sereb},
or whether it is relevant for understanding the $6.5\sigma$ 
discrepancy between the last precise  measurements 
of the neutron lifetime, 
$\tau_{\rm dec} = (885.4 \pm  0.9_{\rm stat} \pm 0.4_{\rm syst})$ s  
\cite{lifetime1} and 
$\tau_{\rm dec} = (878.5 \pm  0.7_{\rm stat} \pm 0.3_{\rm syst})$ s 
\cite{lifetime2}, remain still open: 
in fact, the external magnetic fields were neither screened 
nor controlled in these experiments.  
The fact that the baryon number violating process can be so fast,
much faster the neutron decay, is interesting {\it per se} and 
certainly constitutes a strong challenge.
Such a fast oscillation, with $P(t) = \sin^2(t/\tau_{nn'})$
can occur in a deep cosmos where both ordinary and mirror magnetic
fields are expected to be rather small, and in any case
it would have consequences for the propagation of the
ultra-high energy cosmic rays \cite{BB-nn}.

The effect of the $n-n'$ oscillation can be experimentally tested
by comparing the neutron loss rates for opposite directions
of the applied magnetic field.
By varying the strength of the magnetic field in these experiments
the resonance is achieved when $B=B'$. (recall however 
that resonance conditions change if $n$ and $n'$
states have also different spin-independent potentials.)
It may be convenient to use in the experiments
inhomogeneous magnetic fields with smooth profile
for achieving the MSW-like resonant transitions between $n-n'$.

If the resonant amplification of neutron losses will be really
observed, this would  point to the $n-n'$ oscillation, but
would also allow to provide the crucial test by
observing the neutron regeneration $n \to n' \to n$:
the neutrons disappear from the UCN traps but they
can reappear in the neighboring trap with the same
magnetic conditions with a measurable probability.
In addition, as far as $n-n'$ mixing changes
the neutron precession as well, the effect can be
observed in the neutron precession and depolarization
experiments similar to those that are used for the search
of the neutron EDM.

If the mirror magnetic field has a galactic origin or
it is related to other extended structures like
the mirror molecular clouds or mirror matter in the solar system,
$n-n'$ oscillations should exhibit specific day-night and
perhaps also other seasonal variations.

The non-zero up-down asymmetries (\ref{D-fit}) and (\ref{D-H}),
observed in the  experiments \cite{Ban} and
\cite{Serebrov-New} for the applied magnetic fields
of $B=0.06$ G and $B=0.2$ G respectively,
could be a signal for the $n-n'$ oscillation in the background
of a mirror magnetic field.
However, new dedicated experiments are needed to verify
if at least one of these $3\sigma$ deviations can be real.

\vspace{0.7 cm}

\noindent {\bf Acknowledgements} \\ \\
I am grateful to Anatoly Serebrov for illuminating conversations
concerning the UCN physics,  and to Fabrizio Nesti
and Francesco Villante for many helpful discussions.
I also would like to thank  A. Knecht and A. Mchedlishvili for
clarifying some technical details regarding the experiment \cite{Ban}. 
The work is supported in part by the MIUR grant for the Projects of
National Interest PRIN 2006 "Astroparticle Physics",
and in part by the European FP6 Network "UniverseNet"
MRTN-CT-2006-035863.

\vspace{0.5cm}

\noindent
\underline{\sl Note added}.
I was about to submit this work for publication
but the April 6 Earthquake in L'Aquila made it impossible:
the University building was strongly damaged and all materials
were blocked inside the office.
Only two months later I could enter my office and recover
this manuscript which
I submit after minor cosmetic chan\-ges. 
I am grateful to Eugenio Salom\`e for the hospitality in Rome, 
Gianni Fiorentini for the hospitality in Ferrara, and 
Masud Chaichian and Anca Tureanu for the hospitality 
in Helsinki during the difficult period after the earthquake.  
I dedicate this work to 300 citizens of my town who died 
in the Earthquake.

\end{document}